\documentclass[aps,prl,showpacs,twocolumn,superscriptaddress]{revtex4}

\usepackage{epsfig}
\usepackage{graphicx}
\usepackage{amsmath}
\usepackage{amssymb}
\usepackage{amsfonts}
\usepackage{dcolumn}

\newcommand{\eqb}{\begin{eqnarray}}
\newcommand{\eqe}{\end{eqnarray}}

\begin{document}

\title{Dissipation-induced coherent structures in Bose-Einstein condensates}

\author{Valeriy A. Brazhnyi}

\affiliation{Centro de F\'isica Te\'orica e Computacional, Universidade de Lisboa, Complexo
Interdisciplinar, Avenida Professor Gama Pinto 2, Lisboa 1649-003, Portugal}
 
 \author{Vladimir V. Konotop}
 
 \affiliation{Centro de F\'isica Te\'orica e Computacional, Universidade de Lisboa, Complexo
Interdisciplinar, Avenida Professor Gama Pinto 2, Lisboa 1649-003, Portugal}

\affiliation{Departamento de F\'isica, Faculdade de Ci\^encias, Universidade de Lisboa, Campo
Grande, Ed. C8, Piso 6, Lisboa 1749-016, Portugal }

\author{V\'{\i}ctor M. P\'erez-Garc\'{\i}a}
\affiliation{Departamento de
Matem\'aticas, E. T. S. de Ingenieros Industriales, and Instituto de Matem\'atica Aplicada a la Ciencia y la Ingenier\'{\i}a, Universidad de Castilla-La
Mancha 13071 Ciudad Real, Spain}

\author{Herwig Ott}

\affiliation{Institut f\"ur Physik, Johannes Gutenberg Universit\"at, 55099 Mainz, Germany}

\begin{abstract}

We discuss how to engineer the phase and amplitude of a complex order parameter using localized dissipative perturbations. Our results are applied to generate and control various types of atomic nonlinear matter waves (solitons)
 by means   of  localized dissipative defects.

\end{abstract}
\pacs{05.45.Yv, 03.75.Lm }

\maketitle

\emph{Introduction.-} Dissipation is one of the main forces acting against the formation of nonlinear coherent structures in extended systems. When dissipation is present in systems without  additional gain mechanisms, typically all excitations decay into the regime of linear waves.
In this Letter we discuss how, contradicting  this general principle, a localized dissipation can be used to engineer the phase of  systems governed by complex order parameters through the generation of currents that imprint the required phases in the system. Although our results have general implications we  discuss examples of systems ruled by an universal model of mathematical physics: the nonlinear Schr\"odinger  equation (NLS). As to application fields we will focus on studying the possibility of controlling the phase of Bose-Einstein condensates (BECs) to generate different types of coherent structures.

It was soon after the realization of BECs that their potential to support nonlinear coherent structures was recognized. The list of experimentally found  excitations includes: dark~\cite{dark,dark2} and bright \cite{bright,bright2} solitons, vortices, vortex lattices and related structures \cite{vortices}, vortex rings \cite{vrings},  gap solitons \cite{gap},  shock waves \cite{shockwaves}, different types of vector solitons \cite{vector1} and Faraday waves \cite{Faraday}. Many techniques have been discussed to generate these structures but essentially all of them are ``conservative" in nature   based either on time-varying potentials,  spatially selective optical transitions or on tailored interatomic interactions.

In ultracold quantum gases dissipative mechanisms are related to inelastic collisions \cite{threebody}, interaction with the thermal component \cite{becdi} or collapse dynamics \cite{bright2,dwave}. While the latter can result in surviving matter-waves leading to a ``non-destructive" effect of dissipation \cite{VP} on certain nonlinear structures, dissipation is generally found to damp the excitations and act against the generation or survival of coherent structures.
In this Letter, however, we  show that under certain conditions \emph{a properly localized dissipation}, e.g.  the one recently demonstrated with the help of a focused electron beam \cite{ebeam}, \emph{can be used for generation and control of matter waves}. 

\emph{Physical system and model equations.-} We  consider a quasi-one dimensional (1D) BEC in the mean field limit subject to a localized dissipative defect \cite{ebeam} described phenomenologically by the distribution $\gamma(x)$:
\begin{eqnarray}
\label{NLS_diss}
i \psi_t=- \psi_{xx}+\sigma|\psi|^2\psi -i\gamma(x)\psi.
\end{eqnarray}
Here 
time and space are measured in units of $2/\omega_\bot$ and $a_\bot=\sqrt{\hbar/m\omega_\bot}$  respectively,  where $\omega_\bot$ and $a_\bot$ are the frequency and linear oscillator length of the transverse harmonic trap, and $\sigma=$sign$(a_s)$. The macroscopic wave function is normalized as $\int|\psi|^2dx=4{\cal N}_0|a_s|/a_\bot $,  ${\cal N}_0$ and  $a_s$ being  the initial number of atoms and the scattering length, respectively (see e.g.~\cite{BK}). The applicability of the 1D limit is determined by the condition $a_\bot\ll \xi=(8\pi n_0 |a_s|)^{-1/2}\ll a_\| $, where $a_\| $ is the longitudinal dimension of the condensate, $\xi$ is either the healing length, when excitations against 
a constant 
background of density $n_0$ are under consideration, or a soliton width when one deals with a spatially localized solution (then $n_0$ is  a maximal density of atoms).  The local dissipation \cite{ebeam} is characterized by a temporal scale $\delta\tau$ which is the time interval between subsequent events of eliminating individual atoms from the atomic ensemble. The mean-field approximation  is applicable   if $\delta\tau$  is negligible, i.e. if  $\omega_\bot\delta\tau\ll 1$, what is verified for typical configurations where $\omega_\bot \sim 2\pi \times100$Hz and $\delta\tau\sim 100\,\mu$s.

\emph{Effect of  the localized dissipative defect. -} To understand the physics of the phenomena discussed below
we notice that the rate of particle decay in (\ref{NLS_diss}) is given
by $\gamma(x) n(x)$ where $n$ is the local particle density. Thus, the effect dissipation does not act close to the zeros of $n(x)$ and eventually the loss can become negligible on the scale of the total population of the ground state provided the location of the defect matches the zeroes of the density.

Let us first concentrate on repulsive interatomic interactions ($\sigma=1$). 
If we define $\zeta$ to be the defect size, i.e. the domain where $\gamma(x)$ significantly deviates from zero it is obvious that when $\zeta\to0$ then   $\gamma(x)=\gamma_0\delta(x)$,  $\delta(x)$  being the Dirac delta, and the real constant $\gamma_0$ characterizing the strength of the defect (a dissipative $\delta$-impurity).  In that limit a stationary dark soliton solution of Eq. (\ref{NLS_diss}) has the form $\psi_d=\rho \exp\left(-i\rho^2 t\right)\tanh\left(\rho x/\sqrt{2}\right)$. Obviously, in that limit the dissipative losses, being localized exactly on the zeros of the density, are negligible. 

In the opposite limit where $\gamma(x)\equiv\gamma_0=$const a simple solution of Eq.~(\ref{NLS_diss}) reads
\begin{equation}
\label{exp}
\psi(x,t)=\rho_0 e^{ -\gamma_0 t-i\theta(t)},\quad \theta(t)=
\tfrac{\rho_0^2}{2\gamma_0}\left(1-e^{-2\gamma_0 t}\right)
\end{equation}
where 
$\rho_0 $ is the initial amplitude.

\begin{figure}[ht]
\epsfig{file=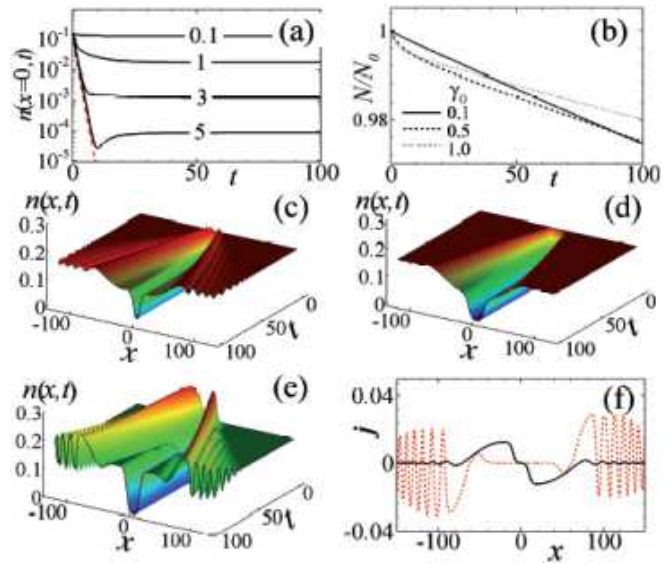,width=\columnwidth} 
\caption{ (a) Evolution of the density for different widths 
of the  defect (\ref{defect}) with $\gamma_0=0.5$.  
The dashed red line shows the exponential decay  (\ref{exp}). 
 (b) Dynamics of $N(t)$ normalized to its initial value $N_0=2\rho_0^2 L$ for different strengths of the  defect. 
In (c) and (d)  surface plots of $n$ for the defect with $\zeta=1<\zeta_{cr}$ and $\zeta=5>\zeta_{cr}$, respectively. In (e) the same as in (d) but for $\gamma_0=-0.5i$ (conservative defect).
(f)  Snapshots of the current density  $j = -|\psi|^2 \cdot [\arg(\psi)]_x$ at $t=100$ for the  defect from (d) (black solid line)  and  for the conservative defect from (e) (red dashed line). 
  In  all panels $\sigma=1$ and $\rho_0=0.4$.
} 
\label{fig_1}
\end{figure}

Let us now consider the more interesting and physically relevant case of a finite size dissipative ``defect"
\begin{eqnarray}
\label{defect}
\gamma(x)=\gamma_0 \exp\left(-x^2/2\zeta^2\right).
\end{eqnarray}
The decay of the atomic density in the center of the defect is characterized by the time $t_\gamma=1/\gamma_0$. The second temporal scale depends on the size of the defect and is given by the time $t_s$ required for sound waves to leave the defect. Since the speed of  sound near the defect core is $c(t)=\sqrt{2}\rho(t)\approx\sqrt{2}\rho_0e^{-\gamma_0t}$ we estimate $t_s=-\log\left[1-\gamma_0\zeta/(2\sqrt{2}\rho_0)\right]$. Thus, if $t_s\ll t_\gamma$, sound waves can emerge from the defect [Fig.~\ref{fig_1}(c)]. However if $t_s\gg t_\gamma$  the sound waves are strongly attenuated and the  density decays smoothly in the vicinity of the defect.  As it is clear this last scenario is always realized for $\zeta\gtrsim\zeta_c$ where $\zeta_{c}=2\sqrt{2}\rho_0/\gamma_0$ [Fig.~\ref{fig_1}(d)].

To confirm these ideas, we have simulated Eq. (\ref{NLS_diss}) with boundary conditions $\psi(\pm L,t) = \rho_0$ for large values of $L$. Typical outcomes are summarized in  Fig.~\ref{fig_1}.  Comparing the evolution of the density at $x=0$ for different widths of the defect [Fig.~\ref{fig_1}(a)] we find that, for sufficiently small $\zeta$, $n(0,t)$ decays monotonically, while non monotonic behavior appears for $\zeta>\zeta_{cr}\approx 2.26$. The dynamics allways ends up in some asymptotic regime, and the differences in the ways of how this regime is achieved are due to the excitation of sound waves, which are clearly seen in the case of the narrow  defect [Fig. \ref{fig_1}(c)], while strongly  attenuated for the case of the wide defect [Fig.~\ref{fig_1}(d)]. We have also simulated the evolution of an initially uniform BEC when the defect is conservative [Fig.~\ref{fig_1}(e)]~\cite{com}. While  the creation of a minimum of the atom density and  the excitation of sound waves propagating outwards the defect look similar to the case of the dissipative defect, the current density reveals significant differences [Fig.~\ref{fig_1}(f)]. In the former case, the currents are transient and due to the kicking of the atoms from the high-potential regions around the defect. However, the dissipative defect  is a sink of energy thus, the currents point towards the defect and are \emph{permanent} in this scenario. \emph{The presence of the dissipative defect implies the imprinting of a nontrivial phase on the condensate wavefunction} [Fig.~\ref{fig_1}(f)].

Fig.~\ref{fig_1}(b) shows another interesting phenomenon: a more intensive defect  results in less intensive loss of condensed atoms at large times. More efficient atomic evaporation by  a weaker defect   can be explained from the fact that such defect created a hole with a finite density leading to the long-term action of the dissipation, while a strong defect results in fast local density decay to zero and thus to suppressed long-term effect of the dissipation.

\begin{figure}[h]
\epsfig{file=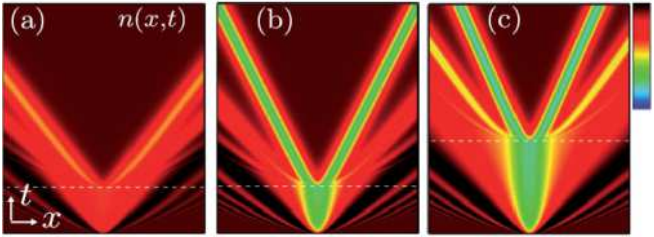,width=\columnwidth} 
\caption{Generation of dark solitons for $\gamma_0=0.5$, $\rho_0=0.4$ and $\sigma=1$. Shown are density plots 
with (a)  $\zeta=0.1$, $t_f=15$; (b) $\zeta=0.5$, $t_f=15$ and (c) $\zeta=0.5$,$t_f=40$. The time range shown is $t\in[0,100]$ and the spatial range $x\in [-50,50]$.
} 
\label{fig_2}
\end{figure}

\emph{Generation of dark solitons.-} 
Next, we discuss the ensuing dynamics after removing the dissipative defect at time $t_f$ (Fig.~\ref{fig_2}).
It can be seen how the ``initial" dark hole splits into a packet of sound waves and  dark solitons propagating outwards the impurity. 
This part of the dynamics is described by the integrable NLS equation allowing for exact computation of the number of solitons
and in this sense is not related to the type of the defect creating the initial intensity hole. The main difference between conservative and  dissipative defects, however, steam form the initial phase shown in panels (c) and (e) of Fig.~\ref{fig_1}: in the last case the excitation of linear  waves (and eventually small amplitude dark solitons) is strongly suppressed, what makes the process more controllable.
A qualitative feature of the phenomenon displayed in Fig.~\ref{fig_1}~(f) is that while the dissipative defect is acting, its main effect is imprinting a phase variation propagating outwards the defect.  
In Fig.~\ref{fig_2} one observes the small amplitude waves generated at $t<t_f$ as dark solitons created by the phase imprinting, while at $t_f$ when the defect is switched off, the shape of the dark hole changes generating two well-separated deep (slow) dark solitons (at this stage, the generation of solitons is dominated by the density engineering). In other words  {\em the dissipative defect provides a controllable way of generating dark solitons based on phase and density engineering.}

\begin{figure}[h]
\epsfig{file=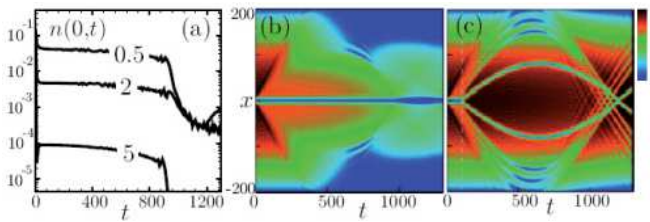,width=\columnwidth} 
\caption{Effect of the dissipative defect on the condensate in a trap for 
$\sigma=1$, $\nu=0.002$, $\rho_0=0.4$, and
$\gamma_0=0.5$.   
 (a) $n(0,t)$ for different defect widths: $\zeta=0.5; 2; 5$.
(b) Pseudocolor plot of the density evolution   for $\zeta=2$.  (c) Same as in (b) but removing dissipation at $t_f=100$. 
} \label{fig_3}
\end{figure}

The situation in a trap might be very different because of a finite number of condensed atoms, resulting in degradation of the whole picture after  significant loss of particles. To study this scenario, we have considered an additional term in Eq.~(\ref{NLS_diss}) of the form  $\nu^2 x^2 \psi$ where $\nu$ determines the linear oscillator frequency of the longitudinal harmonic trap.
In Fig.~\ref{fig_3}(a) we present the evolution of the density in the center of the trap for different widths of the dissipative defect. The initial dynamics (not visible on the scale of the figure) is very similar to that of the case with constant background [Fig.~\ref{fig_1}(a)]. 
With time the loss of particles results in a significant decrease of the  background what takes the system beyond the mean-field regime. 
This happens however after a sufficiently long time: $\approx 800$ in our dimensionless units.
When the dissipative effect is switched off soon enough,  we still observe [Fig.~\ref{fig_3}~(c)] the generation of dark solitons which are later reflected by the parabolic potential and start their characteristic oscillatory motion ~\cite{dark2,BK}.

\begin{figure}[h] 
\epsfig{file=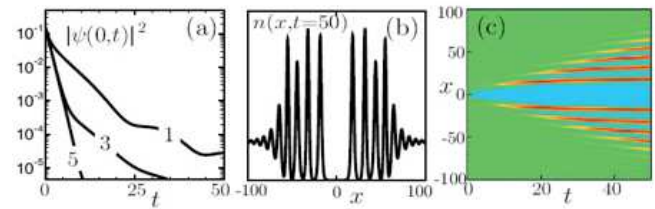,width=\columnwidth} 
\caption{Generation of bright solitons for
$\gamma_0=0.5, \rho_0=0.4$. 
 (a) $n(0,t)$ for different defect widths ($\zeta =1,3,5$).
(b) density profile at $t=50$ (b) and (c) density evolution for $\zeta=3$.
} 
\label{fig_5}
\end{figure}

\emph{Bright solitons.-} When the interatomic interactions are  attractive, $\sigma<0$, a uniform background  is modulationally unstable. In that case, the instability is triggered by  the dissipative defect leading to the emergence of the spatial structures as it is shown in Fig.~\ref{fig_5}. 
In comparison with the repulsive case we observe a faster decay of the amplitude in the defect domain due to the stronger attraction of atoms by the populated regions than by the defect region (in the repulsive case atoms are repelled  from the densely populated domains). The development of the  instability leads to the creation of solitonic pulses [Fig.~\ref{fig_5}~(c)]. This is another counter-intuitive phenomenon since solitons are characteristic excitations of Hamiltonian  conservative systems. However, it is easy to see that after the density hole is created the loss of the atoms is significantly reduced and almost everywhere the system remains to be ``almost-conservative".  

\begin{figure}[ht]
\epsfig{file=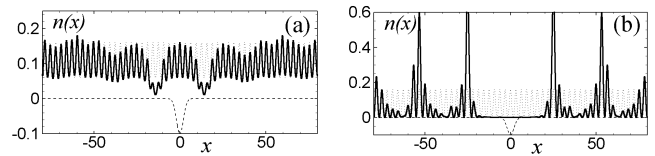,width=\columnwidth}
\caption{Generation of  dark (a) and bright (b) gap solitons after switching off the defect, with $\gamma_0=0.5$ and $\zeta=2$, at $t_f=40$ and $t_f=100$, respectively (solid lines). The initial profiles (dashed lines) are Bloch states for $A=-1$ and $\rho_0=0.5$. The dashed line indicates the  defect profile. } 
\label{fig2_off_OL}
\end{figure}

\emph{Gap solitons.-} Optical lattices (OLs) have proven to be one of the most successful tools for matter-wave management~\cite{BraKon}. Now we  consider again Eq. \eqref{NLS_diss} but with the addition of the periodic potential $V(x)=A\cos(2x)$, whose period  is chosen to be $\pi$ provided the amplitude $A$ is measured in the units of the recoil energy $E_r$. The background solution must be taken as a Bloch state, $\varphi_{\alpha}(x)$ bordering an edge of a gap of the energy spectrum, $\psi(x,0)=\rho_0 \varphi_{\alpha}(x)$ ($\rho_0$ measures the density amplitude and we consider the first lowest gap).   
\begin{figure}[ht]
\epsfig{file=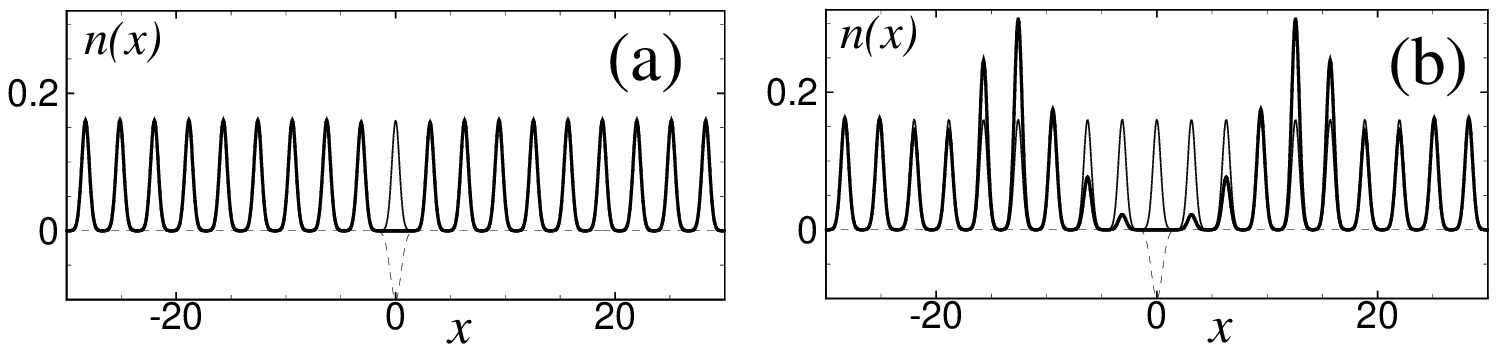,width=\columnwidth} 
\caption{ Initial (thin) and final (thick) at $t=1000$   profiles of the density with, $\sigma=-1$, (a) and  without, $\sigma=0$, (b) the nonlinearity. The parameters are $\rho_0=0.4$, $\zeta=0.5$, and $A=-10$. 
The dashed line illustrates the  defect. 
} \label{fig1_OL_new}
\end{figure} 

We consider $\sigma=-1$ and moderate OL depths ($A \simeq 1$). Then a Bloch state background is stable (unstable) when it borders the  bottom (top) of a band corresponding to positive (negative) effective mass. The behavior of  Bloch states affected by the localized dissipative defect closely resemble the dynamics in a homogeneous condensate, with the major qualitative difference that both dark and bright   solitons can be generated in an attractive BEC, starting with states with positive and negative effective masses, respectively [Fig.~\ref{fig2_off_OL}].


In deep optical lattices $|A|\gtrsim 10$, however, we have arrays of well separated condensates, among which atomic exchange can occur due to the tunneling  and the tight binding approximation becomes more adequate~\cite{BraKon}. The most exciting effect which can be observed in this situation is the creation of persisting density holes due to the suppression of  tunneling by the attractive nonlinearity. Fig.~\ref{fig1_OL_new}, where we compare the dynamics with and without the nonlinearity, illustrates this phenomenon.

\begin{figure}[h] 
\epsfig{file=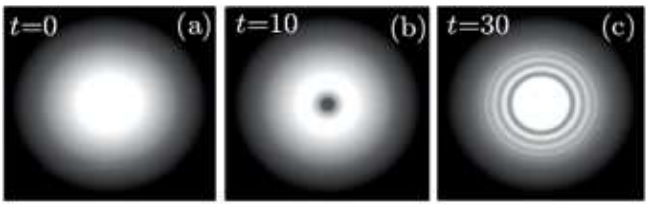,width=\columnwidth} 
\caption{Density plots 
on the region $(x,y) \in [-30,30]\times [-30,30]$ for different times. 
The parameters are $t_f=10$, $\gamma_0=0.5$, $\zeta=2$,  $\rho_0=0.66$, and $\nu=0.01$.} 
\label{fig_7}
\end{figure}
The stationary density hole created in the case $\sigma=-1$ (panel a) is not a dark gap soliton: in our case the initial alternating phase of the Bloch state is preserved and the concept of effective mass loses its significance because the bands are very narrow. Thus the created hole is supported only by the interplay between the tunneling and the attractive nonlinearity and does not have any analogy in the case of a homogeneous medium. The phenomenon can be understood in terms of the effective potential. Indeed, after long enough time the distribution $n(x)$ becomes (quasi-)stationary, even in the presence of the dissipative defect, and $V_{\text{eff}}(x)=V(x)-n(x)$ can be interpreted as the effective potential affecting a single particle. Since for $V \sim 10\,E_r$ the band width is of order of $0.01E_r$ and  $n\sim 0.15\,E_r$,   in the absolute minimum of $V_{\text{eff}} $  the lowest energy levels are shifted down,  the shift being an order of magnitude larger than the energy splitting due to hopping,
resulting in the suppression of tunneling to the central cell.
  
\emph{Multidimensional examples.- } A localized dissipative defect can be used for generating  higher dimensional nonlinear structures.  As an example we have simulated the effect of a 2D Gaussian defect (\ref{defect}) in a repulsive pancake shape BEC  in a harmonic trap $\nu^2 (x^2+y^2)$. After switching off the defect at time $t_f$  we observe the generation of ring dark solitons \cite{RK1}, as shown in Fig.~\ref{fig_7}.

\emph{Conclusion.-} We have shown how a dissipation source can be used to generate nonlinear coherent excitations such as dark, bright, gap or ring dark solitons by engineering  both the phase and amplitude in systems described by NLS equations. Our results can be applied to matter wave management in BECs.
 
V.A.B.   acknowledges support from the FCT grant, PTDC/FIS/64647/2006. V.M.P.-G. is supported by grants FIS2006-04190 (Ministerio de Ciencia e Innovaci\'on, Spain)  and PCI-08-093 (Junta de Comunidades de Castilla-La Mancha).

\end{document}